\documentclass[prc,preprint,nofootinbib]{revtex4}
\usepackage{amssymb}
\usepackage{longtable}
\usepackage{graphicx}
\usepackage{fleqn}
\usepackage{epsfig}
\usepackage[figuresright]{rotating}

\begin{document}

\preprint{}
\author{P. E. Koehler}
\email{koehlerpe@ornl.gov}
\affiliation{Physics Division, Oak Ridge National Laboratory,Oak Ridge, TN 37831}
\author{J. L. Ullmann, T. A. Bredeweg, J. M. O'Donnell, R. Reifarth, R. S.
Rundberg, D. J. Vieira, and J. M. Wouters}
\affiliation{Los Alamos National Laboratory, Los Alamos, NM\ 87545 }
\title{Spin measurements for $^{147}$Sm$+n$ resonances: Further evidence for
non-statistical effects}
\date{\today}

\begin{abstract}
We have determined the spins $J$ of resonances in the $^{147}$Sm($n,\gamma $%
) reaction by measuring multiplicities of $\gamma $-ray cascades following
neutron capture. Using this technique, we were able to determine $J$ values
for all but 14 of the 140 known resonances below $E_{n}=1$ keV, including 41
firm $J$ assignments for resonances whose spins previously were either
unknown or tentative. These new spin assignments, together with previously
determined resonance parameters, allowed us to extract level spacings ($%
D_{0,3}=11.76\pm 0.93$ and $D_{0,4}=11.21\pm 0.85$ eV) and neutron strength
functions ($10^{4}S_{0,3}=4.70\pm 0.91$ and $10^{4}S_{0,4}=4.93\pm 0.92$)
for $J=3$ and 4 resonances, respectively. Furthermore, cumulative numbers of
resonances and cumulative reduced neutron widths as functions of resonance
energy indicate that very few resonances of either spin have been missed
below $E_{n}=700$eV. This conclusion is strengthened by the facts that, over
this energy range, Wigner distributions calculated using these $D_{0}$
values agree with the measured nearest-neighbor level spacings to within the
experimental uncertainties, and that the $\Delta _{3}$ values calculated
from the data also agree with the expected values. Because a non-statistical
effect recently was reported near $E_{n}=350$ eV from an analysis of $^{147}$%
Sm(\textit{n},$\alpha $) data, we divided the data into two regions; $%
0<E_{n}<350$ eV and $350<E_{n}<700$ eV. Using neutron widths from a previous
measurement (corrected for new unresolved doublets identified in this work)
and published techniques for correcting for missed resonances and for
testing whether data are consistent with a Porter-Thomas distribution, we
found that the $\Gamma _{n}^{0}$ distribution for resonances below 350 eV is
consistent with the expected Porter-Thomas distribution. On the other hand,
we found that $\Gamma _{n}^{0}$ data in the $350<E_{n}<700$ eV region are
inconsistent with a Porter-Thomas distribution, but in good agreement with a 
$\chi ^{2}$ distribution having $\nu \geq 2$ \ We discuss possible
explanations for these observed non-statistical effects and their possible
relation to similar effects previously observed in other nuclides.
\end{abstract}

\maketitle

\section{Introduction}

It recently has been shown \cite{Gl2000}\ that (\textit{n},$\alpha $)
cross-section measurements can be very useful for improving calculated
astrophysical rates for reactions involving $\alpha $ particles.
Furthermore, it has been shown \cite{Ko2004}\ that resonance analyses of
such data can be even more useful in improving these rates. This is because
a resonance analysis can eliminate confounding uncertainties and therefore
allow more direct tests of parameters of nuclear models \cite%
{Ra2001,De2002,Ra2003a}\ used to calculate these rates. However, to obtain
the most useful information from a resonance analysis, it is necessary to
know the spins of the resonances. This can be a problem because most of the
nuclides for which (\textit{n},$\alpha $) cross sections are measurable at
resonance energies have non-zero ground-state spins; hence, two spins are
allowed even for low-energy \textit{s}-wave resonances and it can be
difficult or impossible to determine resonance spins using common techniques.

Information contained in the $\gamma $-ray cascades following neutron
capture reactions can, in principle, sometimes be used to determine
resonance spins. For example, in some cases it is expected that the average
number of $\gamma $ rays in the de-excitation cascades between the capturing
states and the ground state will be different for the two \textit{s}-wave
spins. Consider the case of $^{147}$Sm$+n$. Because the ground-state spin of 
$^{147}$Sm is $I^{\pi }=\frac{7}{2}^{-}$, \textit{s}-wave neutrons lead to $%
3^{-}$ and $4^{-}$ resonances in $^{148}$Sm. In a very simple model in which
only dipole transitions can occur, at least three $\gamma $-ray transitions
are required to reach the $0^{+}$ ground state from a $3^{-}$ excited state
whereas a minimum of four transitions are required in the case of a $4^{-}$
state. Hence, in this very simple model, $3^{-}$ resonances will have an
average multiplicity of three and $4^{-}$ resonances an average multiplicity
of four. In reality, the existence of other multipolarities will both
broaden the multiplicity distributions as well as decrease the difference
between average multiplicities for $3^{-}$ and $4^{-}$ resonances \cite%
{Co68,Co2004}. Detector effects also can cause changes in the measured
multiplicity distributions. However, as demonstrated in Ref. \cite{Ge93} the
remaining $\approx 10\%$ difference in average multiplicity for the two
spins still is measurable and independent of resonance energy and was used
to determine spins of 91 $^{147}$Sm$+n$\ resonances below 900 eV.

More recently \cite{Wa2003}, an algorithm that combined Monte Carlo $\gamma $%
-ray cascades predicted by the nuclear statistical model with a Monte Carlo
particle transport code was used to demonstrate that the predicted and
measured multiplicity distributions for a multi-element NaI detector were in
agreement for $3^{-}$ and $4^{-}$ resonances in $^{149}$Sm$+n$. A similar
technique was used to demonstrate good agreement between the measured and
predicted multiplicity spectra for a multi-element BaF$_{2}$ detector \cite%
{Re2002}.

The spin assignments from Ref. \cite{Ge93}\ were used in Ref.\cite{Ko2004}
in an $\mathcal{R}$-matrix analysis of the $^{147}Sm(n,\alpha )$ data of
Ref. \cite{Gl2000} to determine $\alpha $ widths for 104 resonances below
700 eV. The resulting $\Gamma _{\alpha }$ values revealed some surprises
with respect to theoretical expectations. First, the $\alpha $-width
distributions for both $3^{-}$ and $4^{-}$ resonances did not follow the
expected $\chi ^{2}$ distributions. In particular, the $\alpha $-width
distributions were broader than reduced-neutron-width distributions instead
of being intermediate to the distributions for neutrons and $\gamma $ rays.
Second, the ratio of $\alpha $ strength functions for $3^{-}$ to $4^{-}$
resonances was less than $\ $one half of that predicted by theory.
Furthermore, exploratory calculations were not able to find an $\alpha +$%
nucleus potential that could reproduce the observed $\alpha $ strength
functions as well as the strength function ratio. Trying to reduce the $%
\alpha $ strength function ratio to the observed value quickly led to
strength functions which were orders of magnitude larger than measured. Most
surprisingly, the data indicated that there is an abrupt decrease in the $%
\alpha $ strength function ratio for energies above about 300 eV. Such an
abrupt change cannot be reproduced with any optical model of $\alpha $
strength functions.

As pointed out in Ref. \cite{Ko2004}, the $\alpha $-width distributions as
well as the striking decrease in the $3^{-}-4^{-}$ ratio near 300 eV depend
on accurate spin assignments for the resonances, especially above 300 eV. Of
the 104 resonances fitted in Ref. \cite{Ko2004}, 23 resonances (5 below 300
eV) had tentative spin assignments. Therefore, we decided to make a new
measurement of these resonance spins. It was expected that the new Detector
for Advanced Neutron Capture Experiments (DANCE) at the Los Alamos Neutron
Science Center (LANSCE) would make it possible to improve upon the
measurement of Ref. \cite{Ge93} for several reasons. First, the flux at
LANSCE is several orders of magnitude higher, allowing higher precision
measurements even using smaller samples. Second, the DANCE detector has many
more detector segments and a more sophisticated data acquisition system
making more reliable multiplicity measurements possible. Third, the DANCE
detector is made of BaF$_{2}$ rather than NaI as used in Ref. \cite{Ge93}.
This change should lead to reduced backgrounds and improved timing.

\section{Experiment and data reduction}

The experiment was performed using DANCE on flight path 14 at the Manuel
Lujan, Jr. Neutron Scattering Center (MLNSC) at LANSCE \cite{Li90}. DANCE is
a 4$\pi $ array of 160 BaF$_{2}$ crystals positioned 20 m from the neutron
production target. Details of the apparatus \cite{He2001,Re2004}\ and data
acquisition \cite{Wo2006} have been published elsewhere, so only the salient
features will be given herein.

Neutrons are generated at LANSCE via spallation reactions when an 800-MeV
proton beam strikes a tungsten target. The average proton current on target
was 110-120 $\mu $A and the width of the proton pulses was 125 ns. Flight
path 14 views one of the ambient-temperature water moderators at the MLNSC.
The resulting neutron flux peaks near thermal energy and is approximately
proportional to $1$/$E_{n}$ over the range of our measurements.

The samples were placed inside an evacuated flight tube which was surrounded
by a $^{6}$LiH neutron-scattering shield at the center of the DANCE array.
Three samples of metallic samarium, which were enriched to 97.93\% in $%
^{147} $Sm, 1 cm in diameter, and weighed 1.444, 3.208, and 10.410 mg,
respectively were used. The samples were held in the neutron beam by
attaching them to thin Al foils. Sample-out (blank Al backing foil) and
neutron-scattering (C sample) background measurements also were made under
the same conditions.

The neutron flux was monitored using three different sample/detector
combinations downstream of the main sample position: i) a BF$_{3}$ detector,
ii) a fission chamber containing a $^{235}$U sample, and iii) solid state
surface-barrier detectors which recorded tritons and $\alpha $ particles
from the $^{6}$Li($n$,$\alpha $)$^{3}$H reaction occurring in a $^{6}$LiF
sample.

\begin{figure}[tbp]
\includegraphics*[width=85mm,keepaspectratio]{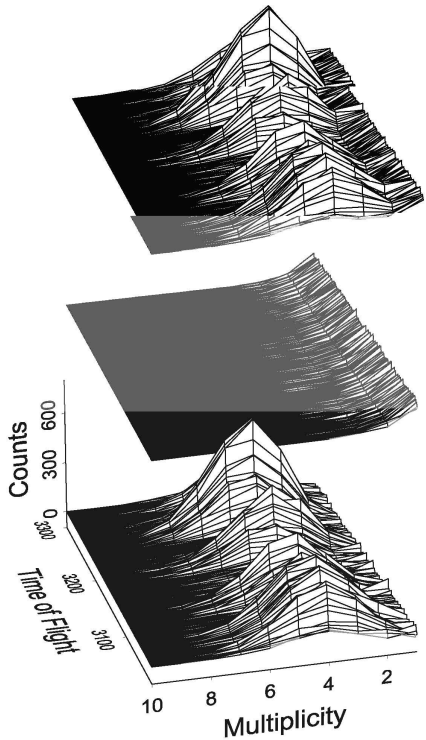}%
\caption{Spectra of counts (arbitrary units) versus multiplicity versus time
of flight for sample-in (top), sample-out (middle), and sample-in minus
sample-out (bottom). The sample-out was normalized to the sample-in spectrum
using the neutron monitor counts. The scales of all three plots are the
same. The neutron energy range of the time-of-flight axes (25 ns/channel) is
roughly 400 to 500 eV.}
\label{2DFig}
\end{figure}

Data were acquired as waveforms, using separate Acqiris transient digitizers
for each detector, over a period of 200 to 250 $\mu $s, triggered by a
timing signal from the accelerator indicating the arrival of a proton pulse
at the neutron production target. Three sets of runs, each with a different
delay for this trigger, were required to cover the entire range from 10 $\mu 
$s before each beam pulse from LANSCE to just below the lowest energy
resonance at 3.397 eV. The waveforms were analyzed in real time to detect
peaks. For each peak, a summary of the peak shape, together with a high
resolution time stamp was written to a disk file. These data were sorted by
a replay routine which generated information such as pulse-height ($\gamma $%
-ray energy), time-of-flight (neutron energy), and cluster multiplicity
(number of $\gamma $ rays detected) for each event. As explained in the
references, cuts were applied to the data to reduce background from
radioactive impurities in the BaF$_{2}$ crystals. In addition, an overall
pulse-height cut on the total $\gamma $-ray energy, $E_{\gamma }=3-8$ MeV,
was used to restrict events to those in the range expected from $^{147}$Sm($%
n $,$\gamma $) reactions. This stage of the analysis resulted in a
two-dimensional spectrum, time-of-flight versus multiplicity, for each of
the runs. The average fluxes recorded by the flux monitors were used to
normalize sample-out runs for background subtraction. Fig. \ref{2DFig} shows
representative sample-in, sample-out, and subtracted two-dimensional spectra.

\begin{figure}[h]
\includegraphics*[width=85mm,keepaspectratio]{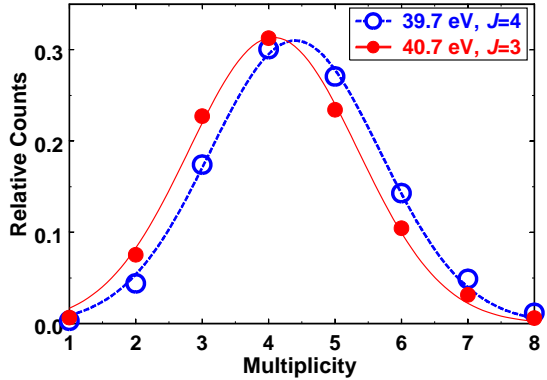}%
\caption{(Color online) Multiplicity spectra for two resonances with firm
spin assignments from previous work \protect\cite{Ge93}. Open and filled
circles are data from our measurements (error bars are smaller than symbol
sizes) and dashed and solid curves are Gaussian fits for the resonances at
39.7 and 40.7 eV, respectively. Fitted mean multiplicities are 4.45 and 4.11
for the 39.7- and 40.7-eV resonances, respectively.}
\label{MultProjFig}
\end{figure}

Projections of the background subtracted spectrum onto the multiplicity axis
for two time-of-flight regions corresponding to resonances having previous
firm spin assignments are shown in Fig. \ref{MultProjFig}.

\begin{figure}[tbp]
\includegraphics*[width=70mm,keepaspectratio]{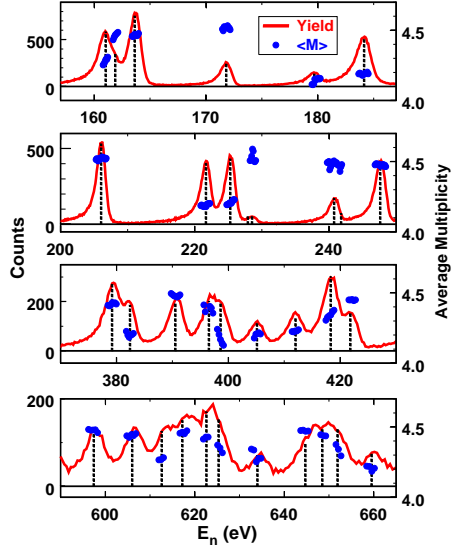}%
\caption{(Color online) Yield (solid red curve, left y axes) and average
multiplicity (solid blue circles, right y axes) versus neutron energy for
three representative energy regions of our data. Dotted vertical lines
indicate positions of resonances identified in previous work. When
resonances are well resolved, they clearly separate into two bands of
average multiplicity. For example, in the top two panels resonances at
163.6, 171.8, 206.03, 240.7, and 247.62 eV have average multiplicities near
4.5 and hence are assigned $J=4$. In contrast, resonances at 179.7, 184.1,
221.65, and 225.28 eV have significantly lower average multiplicities of
about 4.2 and hence $J=3$. However, average multiplicities become less
usefull when resonances are not well resolved. For example, the resonances
at 418.3, 625.3, and 651.9 are only partly resolved from resonances on
either side of them and have average multiplicities half way between the
expected values for the two spin states. As a result, it is not possible to
determine the spins of these resonances using only their average
multiplicities. This situation becomes worse at higher energies.}
\label{AveMvsE}
\end{figure}

These projections verify that there is a measurable, significant difference
in the average multiplicity for the two different \textit{s}-wave resonance
spins. In principle, such projections at each time of flight (or over each
resonance) could be used to determine the average multiplicities and hence
the spins of the resonances as was done in Ref. \cite{Ge93}. This is
demonstrated in Fig. \ref{AveMvsE} where the average multiplicity as a
function of neutron energy is plotted for four energy regions. For this
figure, the average multiplicity is defined by:

\begin{equation}
<M>=\frac{\sum\limits_{i=2}^{9}iY_{i}^{(t)}}{\sum\limits_{i=2}^{9}Y_{i}^{(t)}%
},  \label{Average Multiplicity Equation}
\end{equation}%
where $i$ and $Y_{i}^{(t)}$ are the multiplicity and (background-subtracted)
total yield for that multiplicity, respectively, at neutron energy $E$.
Multiplicities one and greater than nine were not used because the
statistical precision was too poor for these cases. As shown in the top two
panels of Fig. \ref{AveMvsE}, at low energies where most of the resonances
are well resolved, average multiplicities fall into two bands at $<M>\approx
4.2$ and $4.5$ for $J=3$ and $4$, respectively. However, worsening
resolution with increasing neutron energy limits the usefulness of this
approach, and, as shown in the bottom two panels of Fig. \ref{AveMvsE}, once
the resonances are no longer adequately resolved from one another it becomes
difficult or impossible to assign spins using this technique. The problem is
that as instrumental resolution smears the peaks together, the multiplicity
distribution at each neutron energy contains contributions from more than
one resonance. If these resonances have different $J$ values, application of
Eq. \ref{Average Multiplicity Equation} will result in an $<M>$ value
between the values for the two different spins. For example, the resonances
at \cite{Su98} 418.3, 625.3, and 651.9 eV all have $<M>$ values about midway
between the expected values for $J=3$ and $4$. In such cases, the average
multiplicity often will display a positive or negative slope as a function
of neutron energy and, if there is sufficient statistical precision and
there are no other partially-resolved resonances nearby, it may be possible
to discern that the peak in the yield curve actually is due to two
resonances with different spins. For example, the peak near 65 eV was
identified \cite{Ge93} as a doublet, with the lower-energy resonance having $%
J=3$ and the upper one $J=4$, using this technique. On the other hand,
although the $<M>$ versus $E_{n}$ curve displays a slope at the 418.3-,
625.3-, and 651.9-eV resonances, it was not possible to assign firm spins,
or to determine if they were doublets, in any of these cases due to
partially-resolved $J=3$ and $4$ resonances on either side. Another problem
with using $<M>$ to assign spins is that, because it involves division by
the background-subtracted counts, $<M>$ is very noisy between resonances and
near very small resonances where there are few counts. For this reason, $<M>$
is plotted only near the peaks of the resonances in Fig. \ref{AveMvsE}.

To overcome these difficulties, we employed a technique which effectively
uses not only the average multiplicity but also the shapes of the
distributions, and does not require division by the yield. This technique
involves effectively subtracting the prototypical multiplicity distribution
for $J=3$ ($J=4$) resonances from the multiplicity distribution at each
neutron energy, thereby generating a curve as a function of neutron energy
which peaks only at $J=4$ ($J=3$) resonances.

\begin{figure}[tbp]
\includegraphics*[width=70mm,keepaspectratio]{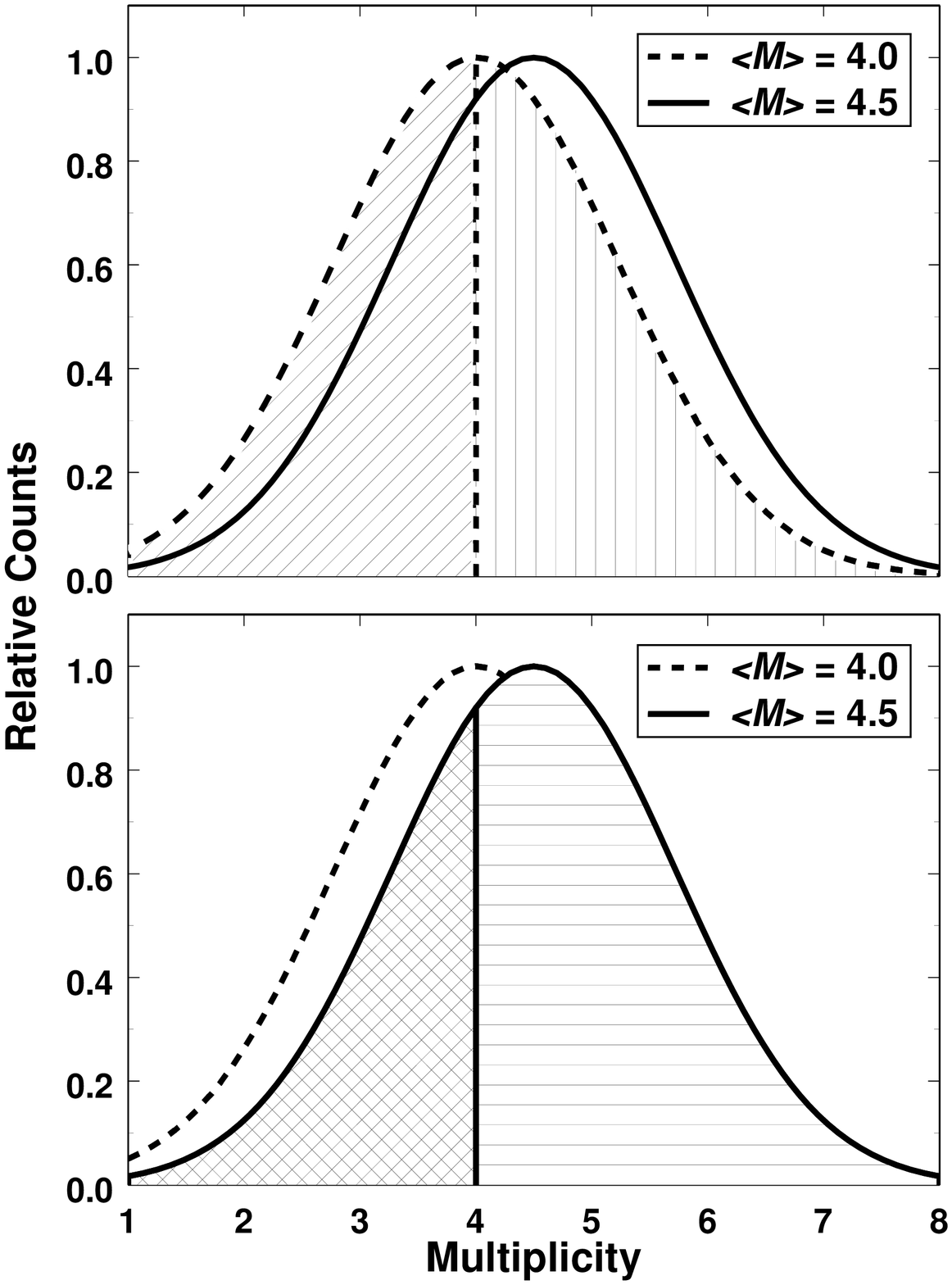}%
\caption{Graphical illustration of Equation \protect\ref{Residual Yield}.
For the purposes of this illustration, it is assumed that multiplicity is a
continuous variable and that multiplicity distributions are symmetric about
their means. Two multiplicity distributions with mean values of $<M>=4.0$
and $4.5$, respectively, are shown. The left and right hatched areas in each
panel represent the two terms in Equation \protect\ref{Residual Yield} for
the two different multiplicity distributions. Integration limits have been
chosen so that they extend for equal ranges of multiplicity on either side
of $M=4$. Under these conditions, the vertically-hatched and
diagonally-hatched areas in the top panel are equal and hence their
difference is zero. However, as shown in the bottom panel, when these same
integration limits are applied to the $<M>=4.5$ distribution, the
horizontally-hatched area is larger than the cross-hatched area. Hence,
subtraction of the latter from the former yields a net positive result.}
\label{MultComboIllus}
\end{figure}

To understand how this technique works, consider that the total yield $%
Y_{i}^{(t)}(E)$ for a given multiplicity $i$ at neutron energy $E$ has, in
general, contributions due to both $J=3$ and $4$ resonances;

\begin{equation}
Y_{i}^{(t)}(E)=Y_{i}^{(3)}(E)+Y_{i}^{(4)}(E).  \label{Total Yield}
\end{equation}%
Assuming that the average multiplicities as well as the shapes of the
multiplicity distributions both remain constant for each of the two spins
(which we have verified for isolated resonances in our data), it is possible
to find a residual yield $Z_{1}^{(3)}(E)$ that will be zero for all $J=3$
resonances;

\begin{equation}
Z_{1}^{(3)}(E)=\sum\limits_{i=a}^{b}Y_{i}^{(3)}(E)-N_{1}\sum%
\limits_{i=c}^{d}Y_{i}^{(3)}(E)=0,  \label{Residual Yield}
\end{equation}%
where $a,b,c,$ and $d$ are integers, and $N_{1}$ is a normalization
constant. For example, if $<M>=4.5$ and the distribution is symmetric, then
Eq. \ref{Residual Yield} is satisfied for $a,b,c,d=5,8,1,4,$ respectively,
and $N_{1}=1$. On the other hand, application of Eq. \ref{Residual Yield} to
a $J=4$ resonance will yield a positive residual because $<M>$ is greater
for $J=4$ resonances than it is for $J=3$. These facts are graphically
illustrated in Fig. \ref{MultComboIllus}. Furthermore, application of Eq. %
\ref{Residual Yield} to the data at energies where the yields contain
contributions from both spins (i.e., Eq. \ref{Total Yield}) will recover the 
$J=4$ component:

\begin{eqnarray}
Z_{1}^{(t)}(E)
&=&\sum\limits_{i=a}^{b}Y_{i}^{(t)}(E)-N_{1}\sum%
\limits_{i=c}^{d}Y_{i}^{(t)}(E)  \label{Residual Yield for Mixed} \\
&=&\sum\limits_{i=a}^{b}[Y_{i}^{(3)}(E)+Y_{i}^{(4)}(E)]-N_{1}\sum%
\limits_{i=c}^{d}[Y_{i}^{(3)}(E)+Y_{i}^{(4)}(E)]  \nonumber \\
&=&\sum\limits_{i=a}^{b}Y_{i}^{(3)}(E)-N_{1}\sum%
\limits_{i=c}^{d}Y_{i}^{(3)}(E)+\sum\limits_{i=a}^{b}Y_{i}^{(4)}(E)-N_{1}%
\sum\limits_{i=c}^{d}Y_{i}^{(4)}(E)  \nonumber \\
&=&\sum\limits_{i=a}^{b}Y_{i}^{(4)}(E)-N_{1}\sum%
\limits_{i=c}^{d}Y_{i}^{(4)}(E),  \nonumber
\end{eqnarray}%
where, in the last step, Eq. \ref{Residual Yield} was used to eliminate the
first two terms in the third line. Similarly, a second residual yield $%
Z_{2}(E)$ can be found that will be zero for all $J=4$ resonances;

\begin{equation}
Z_{2}^{(4)}(E)=\sum\limits_{i=e}^{f}Y_{i}^{(4)}(E)-N_{2}\sum%
\limits_{i=g}^{h}Y_{i}^{(4)}(E)=0.  \label{Residual Yield 2}
\end{equation}

\begin{figure}[tbp]
\includegraphics*[width=70mm,keepaspectratio]{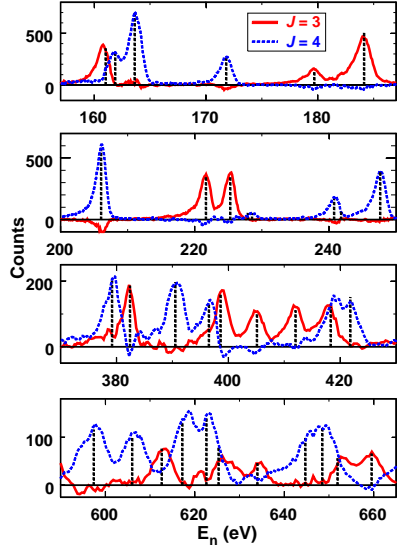}%
\caption{(Color online) Two different linear combinations of multiplicities
versus neutron energy. The red solid curves were calculated using Eq. 
\protect\ref{J=3 Equation} which accentuates $J=3$ resonances. Similarly,
the blue dashed curves were calculated using Eq. \protect\ref{J=4 Equation}
which accentuates $J=4$ resonances. Dotted vertical lines indicate positions
of resonances identified in previous work. The data have been smoothed over
3 to 5 channels to reduce statistical fluctuations.}
\label{MultCombos}
\end{figure}

Because $<M>$ was between $4$ and $5$ for both spins, the summation limits
in Equations \ref{Residual Yield} and \ref{Residual Yield 2} were chosen so
that one sum ended at $i=4$ while the second began at $i=5$. Normalizations $%
N_{1}$ and $N_{2}$ were determined empirically to yield zero net counts in
the vicinity of $J=3$ and $4$ resonances, respectively while yielding net
positive counts for resonances of the other spin. The actual equations used
are given in equations \ref{J=3 Equation} and \ref{J=4 Equation}. Curves
resulting from these equations are shown over the same energy regions as in
Fig. \ref{AveMvsE}, in Fig. \ref{MultCombos}, where the curve labeled $J=3$
was calculated according to:

\begin{equation}
Z_{2}^{(t)}=[0.88\times
\sum\limits_{i=2}^{4}Y_{i}^{(t)}(E)-\sum\limits_{i=5}^{9}Y_{i}^{(t)}(E)]/1.3.
\label{J=3 Equation}
\end{equation}%
Similarly, the curve labeled $J=4$ was calculated using the formula:

\begin{equation}
Z_{1}^{(t)}=\sum\limits_{i=5}^{9}Y_{i}^{(t)}(E)-0.63\times
\sum\limits_{i=2}^{4}Y_{i}^{(t)}(E).  \label{J=4 Equation}
\end{equation}


\begin{figure}[tbp]
\includegraphics*[width=70mm,keepaspectratio]{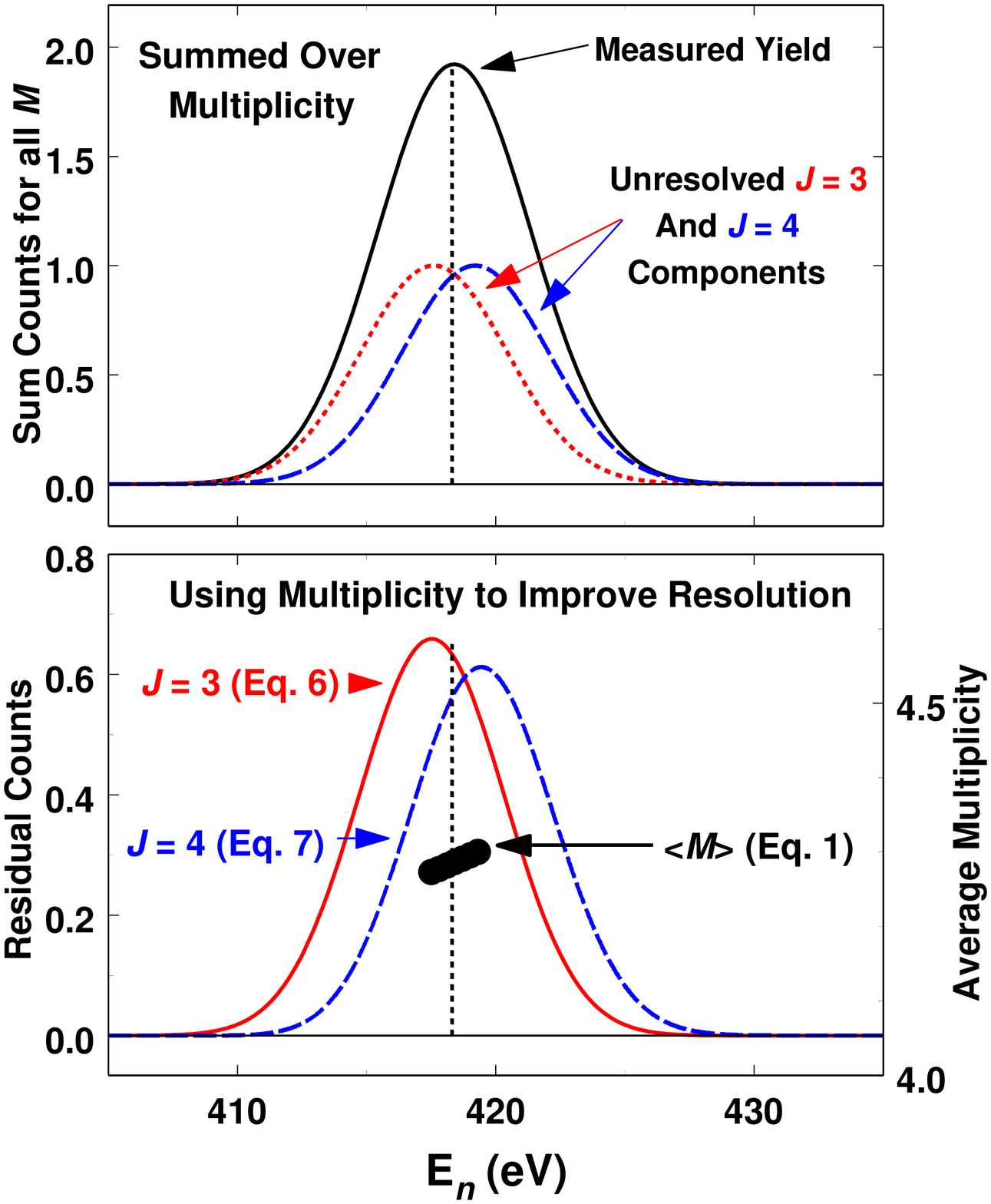}%
\caption{(Color online) Simulations, based on the newly-identified doublet
at 418.3 eV, of our new technique for using multiplicity information to
assign resonance spins. The top panel depicts the simulated measured yield
(black, solid curve) summed over all multiplicities. Using only this
information, it appears that there is a single resonance at this energy. The
other curves in the top panel depict the $J=3$ (short-dashed, red curve) and 
$J=4$ (dashed, blue curve) that were added together to obtain the "Measured
Yield" curve. These two components are of course undetected in the total
yield. Vertical dashed lines in both the top and bottom panels indicate the
position of the previously identified resonance position. The bottom panel
shows the results of using measured multiplicity information. The fitted
multiplicity distributions from Fig. \protect\ref{MultProjFig} were used
together with the individual $J=3$ and $4$ yields in the top panel to
calculate the curves and points in the bottom panel. The solid black circles
represent average multiplicities $<M>$ (right y axis) calculated using
Equation \protect\ref{Average Multiplicity Equation}. The average
multiplicity is about midway between values expected for the two spins and
displays a slight positive slope. These facts hint that this resonance might
be a doublet. However, as shown in Figs \protect\ref{AveMvsE} and \protect
\ref{MultCombos}, in the actual data this slope may be due to the fact that
there is a partially resolved $J=3$ resonance just below and a $J=4$
resonance just above this energy. Hence, it is not possible to draw any firm
conclusions based on $<M>$. The solid red and dashed blue curves in the
bottom panel depict the residual yields (left y axis) calculated using
Equations \protect\ref{J=3 Equation} and \protect\ref{J=4 Equation}, which
reveal both the spins and energies of the individual components of the
doublet.}
\label{JLCIllus}
\end{figure}

The overall normalization constant in Eq. \ref{J=3 Equation} was chosen to
yield peaks of approximately the same height from both equations so that the
results could more easily be compared to one another on the same graph.
Multiplicities one and greater than nine were not used because the
statistical precision was too poor for these cases. The fact that the spin
assignments for isolated resonances from this technique agree with those
from using just the average multiplicities (both from this work as well as
from Ref. \cite{Ge93}) indicates that the multiplicity distributions do
remain reasonably constant. The main advantage of this technique is that it
makes spin assignments possible for several un- and partially-resolved
resonances for which using $<M>$ failed. For example, as discussed above, it
was not possible to make firm spin assignments for the 418.3-, 625.3-, and
651.9-eV resonances using $<M>$. However, as shown in Fig. \ref{MultCombos},
the peak in the yield curve at 418.3 eV, which previously had been given a
tentative $J=(4)$ assignment, is clearly due to two resonances with the
lower-energy one having $J=3$ and the other $J=4$. Fig. \ref{JLCIllus}
depicts simulations based on this doublet in an attempt to further
illustrate this new technique. In addition, Fig. \ref{MultCombos} shows that
the 625.3-, and 651.9-eV resonances have $J=3$. There were many other
similar cases. Overall, of the 140 resonances below $E_{n}=1$ keV, we were
able to make firm $J$ assignments for 33 resonances with no previous $J$
assignments and eight firm $J$ assignments where previously there were only
tentative assignments \cite{Su98}.

Curves calculated using Equations \ref{J=3 Equation} \ and \ref{J=4 Equation}
were used to assign the resonance $J$ values up to $E_{n}=1$ keV listed in
Tables \ref{ResTable}. We stopped at this energy because statistical
analysis indicated that a significant fraction of resonances were beginning
to be missed because of worsening resolution and statistical precision.
Spins from previous measurements also are given in Table \ref{ResTable}.
Only 14 resonances below $E_{n}=1$ keV (nine below 700 eV) remain without
firm $J$ assignments. Only six of our $J$ assignments disagree with those
given in the compilation of Ref. \cite{Su98}. Of these, our $J$ assignments
for the partially-resolved doublet near 65 eV agree with those of the
primary references \cite{Ge93,Al73a} (indicating that perhaps an error was
made in Ref. \cite{Su98} while compiling the data), another two involve
other partially resolved doublets, and the final two previously were only
tentative assignments. Finally, our data indicate that six previously known
resonances (at $E_{n}=$ 140.0, 290.1, 418.3, 513.5, 546.0, and 765 eV)
actually are doublets. For all but the one at 140.0 eV, our data indicate
that the two spin states are about equally strong, so we split the
previously determined $2g\Gamma _{n}$ values equally between the two members
of the doublet. Our data indicate that the $J=3$ component of the doublet at
140.0 eV is about twice as strong as the $J=4$ one, so we split the previous 
$2g\Gamma _{n}$ value by a ratio of 2:1.

\begin{longtable}{ccccccc}%                                                     
\caption{$^{147}$Sm resonance energies and spins\label{ResTable}} \\
\hline\hline
N & $E_{n}$ (eV) & \multicolumn{4}{c}{$J$} \\ 
& & This Work & Ref. \cite{Su98} & Ref. \cite{Ge93} & Ref. \cite{Al73a} & Ref. \cite{Po72} \\ 
\hline
1 & 3.397 & 3 & 3 &  & 3 & 3 \\ 
2 & 18.36 & 4 & 4 & 4 & 4 &  \\ 
3 & 27.16 & 3 & 3 & 3 & 3 & 3 \\ 
4 & 29.76 & 3 & 3 & 3 & 3 & 3 \\ 
5 & 32.14 & 4 & 4 & 4 & 4 &  \\ 
6 & 39.70 & 4 & 4 & 4 & 4 &  \\ 
7 & 40.72 & 3 & 3 & 3 & 3 & 3 \\ 
8 & 49.36 & 4 & 4 & 4 & (4) &  \\ 
9 & 58.09 & 3 & 3 & 3 & 3 &  \\ 
10 & 64.96 & 3 \footnotemark[2] & (4) & 3 \footnotemark[2] & 3 \footnotemark[3] &  \\ 
11 & 65.13 & 4 \footnotemark[2] & (3) & 4 \footnotemark[2] & 4 \footnotemark[3] &  \\ 
12 & 76.15 & 4 & 4 & 4 & 4 &  \\ 
13 & 79.89 & 4 & 4 & 4 & (4) &  \\ 
14 & 83.60 & 3 & 3 & 3 & 3 & 3 \\ 
15 & 94.90 & 3 & &  &  &  \\ 
16 & 99.54 & 4 & 4 & 4 & (4) &  \\ 
17 & 102.69 & 3 & 3 & 3 & 3 & (3) \\ 
18 & 106.93 & 4 & 4 & 4 & (4) &  \\ 
19 & 108.58 & 4 & 4 & 4 &  &  \\ 
20 & 123.71 & 3 & 3 & 3 & 3 & 3 \\ 
21 & 140.00 & (3) \footnotemark[2] & 3 & 3 & 3 &  \\ 
22 & 140.10 & (4 ) \footnotemark[2] &  &  &  &  \\ 
23 & 143.27 & 4 & 4 & 4 &  &  \\ 
24 & 151.54 & 3 & 3 & 3 & 3 &  \\ 
25 & 161.03 & 3 & 3 & 3 &  & 3 \footnotemark[3]  \\ 
26 & 161.88 & 4 & 4 & 4 &  & 3 \footnotemark[3]  \\ 
27 & 163.62 & 4 & 4 & 4 & (4) &  \\ 
28 & 171.80 & 4 & 4 & 4 & (4) &  \\ 
29 & 179.68 & 3 & 3 & 3 &  &  \\ 
30 & 184.14 & 3 & 3 & 3 & 3 & 3 \\ 
31 & 191.07 & 3 & 3 & 3 &  &  \\ 
32 & 193.61 & 4 & 4 & 4 &  &  \\ 
33 & 198.03 & 3 & 3 & 3 &  &  \\ 
34 & 206.03 & 4 & 4 & 4 & (4) &  \\ 
35 & 221.65 & 3 & 3 & 3 &  & 3 \footnotemark[3]  \\ 
36 & 225.28 & 3 & 3 & 3 &  & 3 \footnotemark[3]  \\ 
37 & 227.9 & (4) & & 4\footnotemark[3] &  &  \\ 
38 & 228.53 & 4 & 4 & 4 \footnotemark[3] &  &  \\ 
39 & 240.76 & 4 & 4 & 4 &  &  \\ 
40 & 247.62 & 4 & 4 & 4 &  &  \\ 
41 & 257.13 & 3 \footnotemark[2] & 3 & 3  \footnotemark[3]&  &  \\ 
42 & 258.00 & 4 \footnotemark[2] & 4 & 4  \footnotemark[3]&  &  \\ 
43 & 263.57 & 3 & 3 & 3 &  &  \\ 
44 & 266.26 & 4 & 4 & 4 &  &  \\ 
45 & 270.72 & 3 & 3 & 3 &  &  \\ 
46 & 274.40 & 3 & 3 & 3 &  &  \\ 
47 & 283.28 & 4 & 4 & 4 &  &  \\ 
48 & 290.10 & (4) \footnotemark[2] & (4) & (4) &  &  \\ 
49 & 290.30 & (3) \footnotemark[2] &  & &  &  \\ 
50 & 308.30 & 3 & 3 & 3 &  &  \\ 
51 & 312.06 & 4 & 4 & 4 &  &  \\ 
52 & 321.13 & 3 & 3 & 3 &  &  \\ 
53 & 330.10 & 3 & 3 & 3 &  &  \\ 
54 & 332.1 & 4 & 4 & 4 &  &  \\ 
55 & 340.4 & 4 & 4 & 4 &  &  \\ 
56 & 349.86 & 3 & 3 & 3 &  &  \\ 
57 & 359.32 & 4 & 4 & 4 &  &  \\ 
58 & 362.15 & 4 & 4 & 4 &  &  \\ 
59 & 379.2 & 4 & 4 & 4 &  &  \\ 
60 & 382.4 & 3 & 3 & 3 &  &  \\ 
61 & 390.5 & 4 & 4 & 4 &  &  \\ 
62 & 396.5 & 4 & (4) & (4) &  &  \\ 
63 & 398.6 & 3 & 3 & 3 &  &  \\ 
64 & 405.1 & 3 & 3 & 3 &  &  \\ 
65 & 412.0 & 3 & 3 & 3 &  &  \\ 
66 & 417.6 & 3 \footnotemark[2] & (4) & (4) &  &  \\ 
67 & 419.2 & 4 \footnotemark[2] & &  &  &  \\ 
68 & 421.8 & 4 & 4 & 4 &  &  \\ 
69 & 433.1 & 4 & & 3 \footnotemark[3]  &  &  \\ 
70 & 435.7 & 3 & 3 & 3 \footnotemark[3] &  &  \\ 
71 & 440.2 & 4 & 4 & 4 &  &  \\ 
72 & 446.9 & 3 & 3 & 3 &  &  \\ 
73 & 458.6 & 4 & 4 & 4 &  &  \\ 
74 & 462.9 & 3 & 3 & 3 &  &  \\ 
75 & 476.0 & 4 & 4 & 4 &  &  \\ 
76 & 479.8 & 3 & 3 & 3 &  &  \\ 
77 & 486.4 & 3 & 3 & 3 &  &  \\ 
78 & 496.2 & 4 & 4 & 4 &  &  \\ 
79 & 498.6 & 3 & (3) & (3) &  &  \\ 
80 & 513.5 & (3) \footnotemark[2] & 4 &  4 &  &  \\ 
81 & 515.4 & (4) \footnotemark[2] & &  &  &  \\ 
82 & 528.9 & 4 & 4 & 4 &  &  \\ 
83 & 532.5 & 3 & 3 & 3 &  &  \\ 
84 & 538.1 & 4 & 4 & 4 &  &  \\ 
85 & 546.0 & (3) \footnotemark[2] & (3) & (3) &  &  \\ 
86 & 546.2 & (4) \footnotemark[2] & &  &  &  \\ 
87 & 553.2 & 3 & 3 & 3 &  &  \\ 
88 & 554.5 & 4 & 4 & 4 &  &  \\ 
89 & 559.7 & 3 & 3 & 3 &  &  \\ 
90 & 563.4 & 4 & 4 & 4 &  &  \\ 
91 & 567.6 & 3 &  & &  &  \\ 
92 & 574.3 & 4 & 4 & 4 &  &  \\ 
92 & 580.2 & 3 & 3 & 3 &  &  \\ 
93 & 587.8 & 3 & 3 & 3 &  &  \\ 
94 & 597.4 & 4 & 4 & 4 &  &  \\ 
95 & 606.0 & 4 & 4 & 4 &  &  \\ 
96 & 612.6 & 3 &  & &  &  \\ 
97 & 617.2 & 4 &  (3) & &  &  \\ 
98 & 622.6 & 4 & &  &  &  \\ 
99 & 625.3 & 3 & &  &  &  \\ 
100 & 634.0 & 3 & 3 & 3 &  &  \\ 
101 & 644.7 & 4 & &  &  &  \\ 
102 & 648.5 & 4 & &  &  &  \\ 
103 & 651.9 & 3 & &  &  &  \\ 
104 & 659.5 & 3 & (4) & (4) &  &  \\ 
105 & 668.8 & 4 & 4 & 4 &  &  \\ 
106 & 677.5 & 3 &  & &  &  \\ 
107 & 683.1 & 4 & &  &  &  \\ 
108 & 687.4 & 4 & &  &  &  \\ 
109 & 697.0 & 4 & (4) &  &  &  \\ 
110 & 702 & 3 & &  &  &  \\ 
111 & 714.0 & 3 & 3 & 3 &  &  \\ 
112 & 724 & 3 & &  &  &  \\ 
113 & 729 & 4 & &  &  &  \\ 
114 & 734 & 3 & &  &  &  \\ 
115 & 744.3 & 4 & 4 & 4 &  &  \\ 
116 & 754 & 4 & &  &  &  \\ 
117 & 758 & 3 &  & &  &  \\ 
118 & 764 & 4 \footnotemark[2] & &  &  &  \\ 
119 & 766 & 3 \footnotemark[2] &  & &  &  \\ 
120 & 796.2 & 3 & 3 & 3 &  &  \\ 
121 & 808.0 & 4 & 4 & 4 &  &  \\ 
122 & 821.0 & 4 & 4 & 4 &  &  \\ 
123 & 836.1 & (4) & 4 & 4 &  &  \\ 
124 & 847 & 4 &  & &  &  \\ 
125 & 850 & (3) &  & &  &  \\ 
126 & 854 & (4) &  & &  &  \\ 
127 & 858 & 4 &  & &  &  \\ 
128 & 864 & 3 &  & &  &  \\ 
129 & 875.2 & 3 & 4 & 4 &  &  \\ 
130 & 880 & 4 &  & &  &  \\ 
131 & 896.1 & (4) & 4 & 4 &  &  \\ 
132 & 911 & 3 & &  &  &  \\ 
133 & 922 & 4 & &  &  &  \\ 
134 & 930 & 3 & &  &  &  \\ 
135 & 935 & 4 &  & &  &  \\ 
136 & 943 & 4 & &  &  &  \\ 
137 & 953 & (3) &  & &  &  \\ 
138 & 962 & 3 & &  &  &  \\ 
139 & 984 & 3 &  & &  &  \\ 
140 & 991 & 4 & &  &  &  \\ 
\hline\hline                                                                   
\footnotetext[1]{Energies from Ref.  \cite{Su98} except for some unresolved doublets.}
\footnotetext[2]{Partially resolved doublet.}
\footnotetext[3]{Unresolved doublet.}
\end{longtable}

\section{Resonance parameter analysis and discussion}

As a result of our new data, almost all the resonances below 700 eV have
firm spin assignments. Therefore, it should be possible to perform a much
better analysis of the resonance parameters than previously was possible.

\subsection{Level spacings and neutron strength functions}

\begin{figure}[tbp]
\includegraphics*[width=85mm,keepaspectratio]{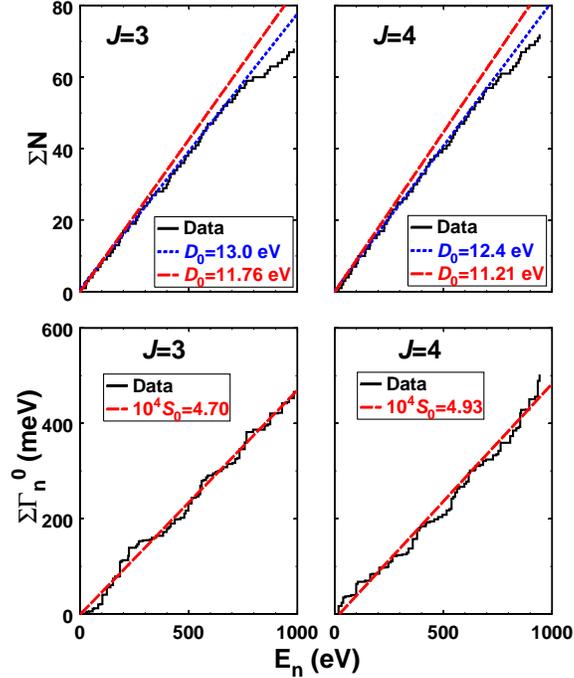}%
\caption{(Color online) Cumulative number of resonances (top) and reduced
neutron widths (bottom) versus resonance energy for $J=3$ (left) and $4$
(right) resonances. Data from measurements are represented by staircase
plots. Short-dashed blue lines in the top panels are linear fits to the data
below 700 eV from which the indicated level spacing values were obtained.
These same level spacing values were used to calculate the Wigner
distributions depicted by dashed curves in Fig. \protect\ref{ELT700Spacings}%
. Long-dashed red lines in the top panels depict level spacings after a
correction for missed resonances was applied. See text for details.
Long-dashed red lines in the bottom panels are fits to the data over the
entire range shown from which the indicated neutron strength functions were
determined.}
\label{D0AndS0}
\end{figure}

Plots of the cumulative number of resonances as a function of resonance
energy are shown in the top part of Fig. \ref{D0AndS0}. Average level
spacings can be calculated from the reciprocals of the slopes of these plots 
\cite{Mu81}. These data indicate that a significant fraction of resonances
are beginning to be missed for energies in excess of 700 eV. Therefore, only
the data below this energy were used to determine the average level
spacings. Dashed lines depict the results of linear fits to the data for $%
E_{n}<$ 700 eV from which average level spacings of $D_{0,3}=12.99\pm 0.93$
eV and $D_{0,4}=12.38\pm 0.85$ eV for $J=3$ and $4$ resonances, respectively
were determined. Uncertainties were calculated according to Ref. \cite{Mu81}%
. The nearly equal level spacings for the two spin groups is in agreement
with Fermi gas model predictions (see, for example Ref. \cite{Ki75}).

Plotted in the bottom part of Fig. \ref{D0AndS0} are cumulative reduced
neutron widths as functions of resonance energy. Neutron strength functions
can be determined from the slopes of these plots \cite{Mu81}. Neutron widths
(except as noted above) were taken from Ref. \cite{Su98}, which is based on
Ref. \cite{Mi81}. Because the measurement technique of Ref. \cite{Mi81} is
expected to miss only resonances having very small neutron widths, and
because such resonances contribute very little to the cumulative reduced
neutron widths, the data over the entire region to 1 keV were used to
determine the neutron strength functions. Dashed lines indicate the results
of straight-line fits to the data from which strength functions $%
10^{4}S_{0,3}=4.70\pm 0.91$ and $10^{4}S_{0,4}=4.93\pm 0.92$ for $J=3$ and 4
resonances, respectively were determined. Uncertainties were calculated
according to Ref. \cite{Mu81}.

\begin{figure}[tbp]
\includegraphics*[width=85mm,keepaspectratio]{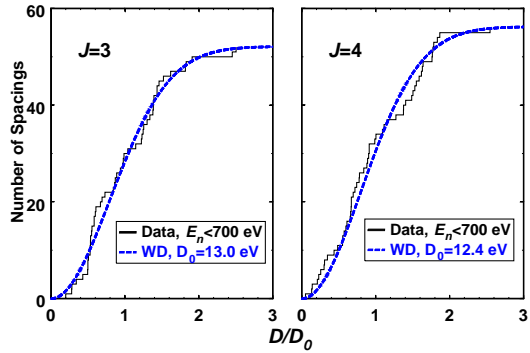}%
\caption{(Color online) Cumulative nearest-neighboor level spacing
distributions for $J=3$ (left) and $4$ (right) resonances. Plotted are the
cumulative number of spacings up to a given value versus that value. The
spacings, $D$, have been normalized to the indicated average spacings $D_{0}$%
. Data from measurements are represented by staircase plots. Dashed blue
curves indicate the expected Wigner distributions.}
\label{ELT700Spacings}
\end{figure}

Further evidence that very few resonances have been missed below 700 eV is
provided by the resonance spacing distributions. The integral
nearest-neighboor spacing distributions for resonances below this energy are
plotted in Fig. \ref{ELT700Spacings}. We plotted integral rather than
differential distributions for these data, as well as for the width
distributions shown below, to avoid possible systematic effects due to the
choice of binning widths. From these plots it can be seen that the measured
spacings are in good agreement with the expected Wigner distributions \cite%
{Ly68}. Furthermore, $\Delta _{3}$ values \cite{Dy63} (which are sensitive
measures of the expected longer range correlations in the level spacings)
calculated from the data (0.40 for both spin states for resonances below 700
eV) are in excellent agreement with the expected values ($0.40\pm 0.11$ for
both spin states). All these results indicate that there are very few
missing or missasigned resonances for $E_{n}<$ 700 eV.

\subsection{Neutron width distributions}

Reduced neutron widths for a single $J$ value are expected to follow a $\chi
^{2}$ distribution with one degree of freedom ($\nu =1$) - the so-called
Porter-Thomas (PT) distribution \cite{Po56}. A $\chi ^{2}$ distribution with 
$\nu $ degrees of freedom of widths $\Gamma $ has the form:

\begin{equation}
P(x,\nu )=\frac{\nu }{2G(\nu /2)}(\frac{\nu x}{2})^{\nu /2-1}\exp (-\frac{%
\nu x}{2})  \label{Chi2DistEq}
\end{equation}

where $P(x,\nu )$ is the probability, $x=\frac{\Gamma }{\langle \Gamma
\rangle }$, $\langle \Gamma \rangle $ is the average width, and $G(\nu /2)$
is the gamma function for $\nu /2$.

The PT distribution has been compared to reduced-neutron-width data in
several instances (e.g., Refs. \cite{Po56,Ca76a,Ca94}) and now is considered
to be a well established fact. However, there are three main problems with
such comparisons. First, the relatively small number of available resonances
limits the statistical precision. Hence, these tests usually employ a
statistical technique such as the maximum likelihood method to determine the 
$\nu $ value of the distribution from the data. Also, the formalism of error
propagation was used in Ref. \cite{Ha84} to derive the standard deviation in
the $\nu $\ value determined from the data given the number of resonances
used. Second, it is an unfortunate fact that the PT\ distribution is
weighted towards small widths that are the most difficult to observe in
experiments. Furthermore, the region of small widths is where the PT
distribution differs most from the next closest $\chi ^{2}$ distribution
having $\nu =2$. Therefore, tests of the PT distribution must include a
consideration of missed resonances. For example, in Fig. 2 of Ref. \cite%
{Po56} several curves are given for different experimental sensitivities, to
be used in determining the $\nu $ value from a set of measured reduced
neutron widths. Third, care must be taken to avoid contamination from $p$%
-wave resonances. Because neutron widths for $p$-wave resonances are, on
average, much smaller than for $s$-wave ones, inclusion of only a small
number of $p$-wave resonances can lead to an erroneously small $\nu $ value
being extracted from the distribution.

As a test case for the PT\ distribution, $^{147}$Sm has the advantages that
a relatively large number of resonances are available and that the data
should be free of $p$-wave contamination. A minimum of 54 resonances were
used in the tests described below, which is more than used in eight of the
fourteen cases studied in Refs. \cite{Ca76a,Ca94}. Furthermore, $^{147}$Sm
is near both the maximum of the $s$-wave as well as the minimum of the $p$%
-wave neutron strength functions ($S_{0}/S_{1}\approx 10$). In addition, due
to its relatively small average level spacing, a sufficient number of $s$%
-wave resonances can be observed at relatively low energies, before the
largest $p$-wave neutron widths become comparable to the smallest $s$-wave
ones. In contrast, many of the nuclides studied in Ref. \cite{Ca94} are near
the peak of the $p$-wave strength function, having $S_{0}/S_{1}\approx 0.4-3$%
, and have level spacings 2.6 to 16 larger than $^{147}$Sm. Therefore, for
these nuclides it was necessary to include resonances to much higher
energies to obtain adequate sample sizes, and to use relatively high
threshold $\Gamma _{n}^{0}$ values to avoid $p$-wave contamination. Because
theoretical distributions for different $\nu $ values differ most at small $%
\Gamma _{n}^{0}$, using a higher threshold limits the sensitivity of the
test.

\begin{figure}[tbp]
\includegraphics*[width=85mm,keepaspectratio]{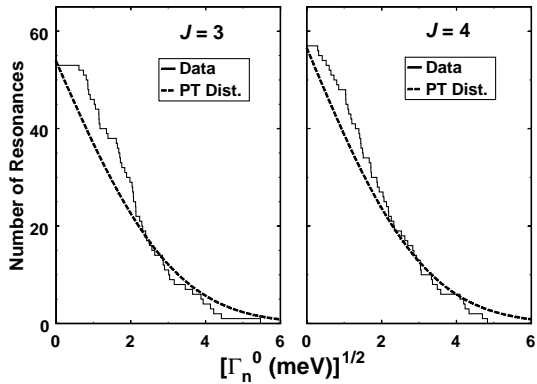}%
\caption{Cumulative distributions of reduced neutron widths for $J=3$ (left)
and $4$ (right) resonances below 700 eV. Plotted are the number of
resonances having a reduced neutron width greater than a given value versus
the square root of that value. Data from measurements are represented by
staircase plots. Dashed curves represent the expected Porter-Thomas
distributions and were calculated using the level spacings and neutron
strength functions determined from the data in Fig. \protect\ref{D0AndS0}.}
\label{NeutWidthDists}
\end{figure}

Given the measured level spacing and strength function (which determine $%
\langle \Gamma \rangle $ and the overall normalization) there are, in
principle, no free parameters when comparing the measured reduced neutron
widths to the expected PT\ distribution. Because we have determined level
spacings and strength functions for both \textit{s}-wave spin states, we can
compare the $\Gamma _{n}^{0}$ distributions for each to the expected PT
distributions as shown in Fig. \ref{NeutWidthDists}. As can be seen in this
figure, there appears to be substantial disagreement between the data and
the expected distributions. To quantify these differences, we used the $%
\Gamma _{n}^{0}$ values together with Eq. 2 and Fig. 2 of Ref. \cite{Po56}
(which are based on the maximum likelihood method) to estimate $\nu $
values. For $J=3$ and 4, the first term on the left hand side of Eq. 2 ($%
\frac{1}{N}\sum \ln (\Gamma _{n,i}^{0}/\langle \Gamma _{n}^{0}\rangle )$,
where the sum runs from $i=1$ to $N$, the number of resonances) in Ref. \cite%
{Po56} equals $-0.50$ and $-0.68$, respectively. To use these values with
Fig. 2 of this reference, it is necessary to choose a threshold value for
the experiment, $x_{\frac{1}{2}}$, which is the antilog of the value of $%
\Gamma _{n}^{0}/\langle \Gamma _{n}^{0}\rangle $ at which the overall
efficiency of detecting a reduced neutron width this small is $\frac{1}{2}$.
According to Ref. \cite{Po56}, the most probable value is $x_{\frac{1}{2}%
}=0.01$, so we used the curve for this value to obtain $\nu =2.0\pm 0.22$
and $1.5\pm 0.22$ for $J=3$ and 4, respectively. The uncertainties were
calculated according to Eq. 2.14 in Ref. \cite{Ha84} from which it can be
concluded that the $\Gamma _{n}^{0}$ distributions for $J=3$ and 4 are 4.5
and 2.3 standard deviations different from the expected value of $\nu =1$
for a PT distribution.

\begin{figure}[tbp]
\includegraphics*[width=85mm,keepaspectratio]{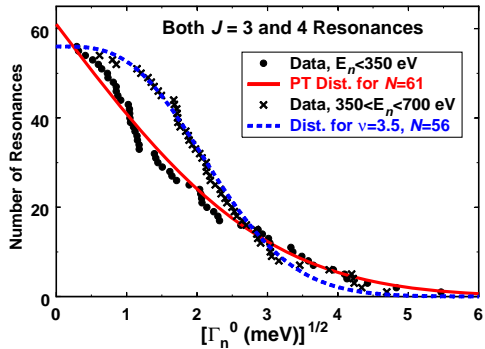}%
\caption{(Color online) Distributions of reduced neutron widths for two
different energy regions. Plotted are the number of resonances (both $J=3$
and 4 combined) having a reduced neutron width greater than a given value
versus the square root of that value. Resonances with $E_{n}<350$ eV and $%
350<E_{n}<700$ eV are shown as solid circles and X's, respectively. We used
symbols rather than the more typical staircase plots for the data so that
they could be distiguished more easily from each other and from the
theoretical curves. The solid red and dashed blue curves are the expected PT
and $\protect\nu =3.5$ distributions, respectively, after corrections for
missed resonances as explained in the text.}
\label{Chi2BothJs}
\end{figure}

Other methods have been devised to correct for missed resonances, and other
statistical tests may be used to ascertain if the data are consistent with a
PT\ distribution. Before proceeding further however, first let us consider
the fact that a non-statistical effect recently was reported \cite{Ko2004}
near $E_{n}=350$ eV from an analysis of $^{147}Sm(n,\alpha )$ data. With
this in mind, we divided the $\Gamma _{n}^{0}$ data into two groups from $%
E_{n}=0-350$ eV and $E_{n}=350-700$ eV. Also, because our analysis indicates
that the average reduced neutron widths are equal for $J=3$ and 4, we
combined the data (as $\Gamma _{n}^{0}$) for the two spins to increase the
statistical precision. In Ref. \cite{Mi81}, the data were combined as $%
g\Gamma _{n}^{0}$ (where $g=\frac{2J+1}{2(2I+1)}$ where $I=\frac{7}{2}$ the
spin of the target nuclide $^{147}$Sm) as typically is done when the
resonance spins are unknown. However, combining two spin groups in this way
implicitly assumes that the number of resonances are proportional to $2J+1$,
which we have shown is not the case. Neutron width distributions for the two
energy regions are shown in Fig. \ref{Chi2BothJs}. From this figure, it
appears that the $\Gamma _{n}^{0}$ distribution changes shape between the
two energy regions. Below 350 eV, the shape appears to be very well
described by a PT\ distribution. Using Eq. 2 and (the $x_{\frac{1}{2}}=0.01$
curve in) Fig. 2 in Ref. \cite{Po56}, leads to $\nu =1.02\pm 0.22$ for the
lower-energy region, in excellent agreement with PT. In contrast, this same
method leads to $\nu =3.5\pm 0.22$ for the $E_{n}=350-700$ eV region, or
more than 11 standard deviations different from $\nu =1$. To obtain this
result, we used the equations in Ref. \cite{Po56} to extend the curves in
Fig. 2 of that reference (which ends at $\nu =2$). For such large $\nu $
values, curves for the different $x_{\frac{1}{2}}$ values are nearly the
same. One problem with the technique of Ref. \cite{Po56} is that the
correction for missed resonances is made using an energy-independent
threshold value $x_{\frac{1}{2}}$, whereas in most experiments the
sensitivity decreases with increasing energy. Therefore, it seems prudent to
employ a more realistic correction for the number of missed resonances.

In Ref. \cite{Fu65}, a technique for calculating the number of missed
resonances was devised which is based on realistic experimental conditions.
The technique as it is laid out in Ref. \cite{Fu65} also assumes the reduced
neutron widths obey a PT\ distribution. We have shown above that the neutron
widths for resonances below 350 eV are in good agreement with PT. Therefore,
we applied the technique of Ref. \cite{Fu65} to the data in this region to
obtain corrected $D_{0}$ and $S_{0}$ values (and hence corrected values for
the number of resonances in the 350-eV interval $N_{corr}$ and corrected
values for the average reduced neutron width) and assumed these values
remain the same for the next 350-eV interval.

\begin{figure}[tbp]
\includegraphics*[width=85mm,keepaspectratio]{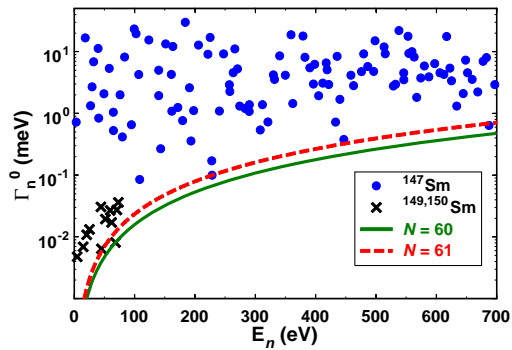}%
\caption{(Color online) Reduced neutron widths for $^{147}$Sm resonances
(blue solid circles) and effective $\Gamma _{n}^{0}$ values for $^{149,150}$%
Sm resonances (black X's) in our DANCE data as functions of resonance
energy. Also shown are threshold curves calculated according to Ref. 
\protect\cite{Fu65} for four ($N=60$, solid geen curve) and five ($N=61$,
long-dashed red curve) missed resonances by $E_{n}=350$ eV. See text for
details.}
\label{Gn0vsE}
\end{figure}

To apply this technique, it is necessary to determine an energy-dependent
threshold $\Gamma _{n}^{0}$ below which resonances are missed, $\delta
(E)=c\langle \Gamma _{n}^{0}\rangle E^{b}$, where $c$ and $b$ are constants
determined from the data and type of experiment, respectively. Ref. \cite%
{Fu65} indicates that $b=1.75$ for the present experiments and, as can be
seen in Fig. \ref{Gn0vsE}, this choice of $b$ seems to agree well with the
experimental threshold across a wide energy range. An examination of the
reduced neutron widths below 350 eV indicates that the most sensitive limit
is set by the 228.53-eV resonance, from which $c^{(1)}=1.22\times 10^{-6}$
is obtained. Following the iterative procedure of Ref. \cite{Fu65}, these
values of $b$ and $c^{(1)}$ lead to a corrected average level spacing of $%
D_{0}=5.74\pm 0.40$ eV (for both spins combined, with uncertainty calculated
according to Ref. \cite{Mu81}) and negligible change to $S_{0}$. Assuming
the relative number of resonances for the two different spins remains
unchanged, the corrected spin-separated average level spacings are $%
D_{0,3}=11.76\pm 0.93$ eV and $D_{0,4}=11.21\pm 0.85$ eV. Hence, this
technique indicates that 5 resonances were missed by 350 eV, or $N_{corr}=61$%
. Peaks in our data due to small amounts of $^{149,150}$Sm (0.50 and 0.17
at\%, respectively) in the sample indicate that this is a conservative
estimate and that the actual number of missed resonances is smaller. Of the
observed $^{149,150}$Sm resonances, the one at 68.3 eV yields the most
sensitive limit. Using this resonance, the parameters in Ref. \cite{Su98},
the assayed amount of $^{149}$Sm in the sample, and the methods of Ref. \cite%
{Fu65} yields $N_{corr}=60$ by 350 eV. Reduced neutron widths for $^{147}$Sm
and effective $\Gamma _{n}^{0}$ values for $^{149,150}$Sm are shown together
with the $N_{corr}=60$ and $61$ threshold curves in Fig. \ref{Gn0vsE}.

In addition to providing corrections for the number of missed resonances,
the calculations above also make it possible to do a more careful and
realistic maximum likelihood analysis as described in Ref. \cite{Ca94}.
Instead of the somewhat arbitrary threshold used in Ref. \cite{Po56}, in the
technique of Ref. \cite{Ca94}, an energy-independent threshold is determined
from the data by examining a plot such as Fig. \ref{Gn0vsE}. The threshold $%
\Gamma _{n}^{0}$ value is chosen such that, within the energy range being
considered, all \textit{s}-wave resonances appear to have been observed and
all \textit{p}-wave resonances excluded. As explained above, the latter
consideration can be neglected in the present case. From Fig. \ref{Gn0vsE},
it can be seen that the conservative ($N=61$) threshold curve implies that $%
\Gamma _{n}^{0}=0.2$ meV is a reasonable threshold value for $E_{n}<350$ eV.
Similarly, $\Gamma _{n}^{0}=0.7$ meV is a reasonable threshold value for $%
E_{n}<700$ eV. With these threshold choices, applying the technique of Ref. 
\cite{Ca94} leads to $\nu =0.91\pm 0.32$ for the $E_{n}<350$ eV region, and $%
\nu =3.19\pm 0.83$ for the $350<E_{n}<700$ eV region. Hence, this improved
analysis leads to the same conclusion as applying the method of Ref. \cite%
{Po56}: The data in the lower energy region are consistent with a PT
distribution, but the higher-energy data are inconsistent with PT. Even if
the very conservative threshold of $\Gamma _{n}^{0}=2.0$ meV is assumed for
the $350<E_{n}<700$ eV region, the $\nu $ value obtained ($2.68\pm 0.76$)
still is inconsistent with a PT distribution at the $2.2\sigma $ level.
Uncertainties were dominated by finite sampling errors, which were
determined in the usual way when maximum likelihood estimators are used, as
described in Ref. \cite{Ca94}. These uncertainties tend to be substantially
larger than those calculated following Ref. \cite{Ha84}, which is based on
the formalism of error propagation.

As a further check, a second statistical technique was applied. The
Kolmogorov-Smirnov (KS)\ test \cite{Co80} can be used to test the hypothesis
that theoretical and measured distributions are equivalent. This test
involves calculating the maximum vertical distance $D^{+}$ between the data
and the hypothesized distribution and accounts for the fact that a limited
number of samples were measured in the experiment. The expected PT\
distribution using the more conservative correction for missed resonances ($%
N_{corr}=61$) is shown in Fig. \ref{Chi2BothJs}. It appears to be in
excellent agreement with the data for $E_{n}<350$ eV and significantly
different from the data for $350<E_{n}<700$ eV. We applied the KS test to
the data in both energy regions. Using $N_{corr}=61$, we calculated $%
D^{+}=0.0919$ and $0.2432$ for the $E_{n}<350$ eV and $350<E_{n}<700$ eV
regions, respectively. These $D^{+}$ values together with the number of
observed resonances were used to calculate $P$ values of $63.40\%$ and $%
99.87\%$ for the $E_{n}<350$ eV and $350<E_{n}<700$ eV regions,
respectively. These $P$ values indicate the hypothesis that the data are
consistent with a PT distribution is accepted for the lower energy region,
but rejected at the $99.87\%$ confidence level for the $350<E_{n}<700$ eV
region. KS tests of these same data compared to a $\chi ^{2}$ distribution
with 3.5 degrees of freedom result in the opposite conclusion; the
hypothesis that the data are consistent with this distribution is accepted
for the higher energy region ($P=40.75\%$), but rejected at the $100.00\%$
confidence level for the $E_{n}<350$ eV region. Although the data in the $%
350<E_{n}<700$ eV region  are in better agreement with larger $\nu $ values,
intermediate degrees of freedom (e.g., $D^{+}=0.1167$ and $P=78.72\%$, for $%
\nu =2$) cannot be excluded. Taken together, both the maximum likelihood and
KS methods indicate the shape of the $\Gamma _{n}^{0}$ distribution changes
from PT to $\nu \geq 2$ at $E_{n}\approx 350$ eV. Results from KS tests of
the various distributions are summarized in Table \ref{KSTestsTable}.

In doing the above tests, we have calculated the correction for missed
resonances using the data in the $E_{n}=0-350$ eV region, and assumed the
same number of resonances ($N_{corr}=61$) in the $350<E_{n}<700$ eV region.
Although it could be argued that it might be better to use the data in the $%
350<E_{n}<700$ eV region to obtain the corrected number of resonances in
this region, there are at least three reasons why our approach is better.
First, as shown in Fig. \ref{Gn0vsE}, sensitivity to small resonances is
greatest at lower energies. Hence by using the data in the $E_{n}=0-350$ eV
region, the correction factor is, in principle, smaller and any unknown
systematic errors should be less important. Second, all such correction
methods must assume a neutron-width distribution. As discussed above,
applying statistical tests to the data in the $E_{n}=0-350$ eV region
indicate that these data are in good agreement with a PT distribution.
Hence, it should be safe to apply the method of Ref. \cite{Fu65} (which
assume a PT\ distribution) to the data in this region to obtain the
corrected number of resonances. On the other hand, these same statistical
tests indicate that the data in the $350<E_{n}<700$ eV region do not follow
a PT\ distribution, so it may not be valid to apply the technique of Ref. 
\cite{Fu65} to obtain the corrected number of resonances in this region from
these data; furthermore, to do so would result in a somewhat circular test
(i.e., assuming $\nu =1$ to obtain the corrected number of resonances with
which to test if $\nu =1$). Third, all such correction techniques are
multiplicative in nature; they obtain the corrected number of resonances by
multiplying the observed number of resonances by a correction factor \
Therefore, a significant systematic error can result if the wrong
neutron-width distribution is assumed. This is because there are fewer
resonances having small neutron widths for a $\nu =3.5$ distribution than
for a PT one. Therefore, for a given threshold such as shown in Fig. \ref%
{Gn0vsE}, fewer resonance will be missed for a $\nu =3.5$ distribution than
for a PT one. Hence, if a PT distribution is assumed, but the distribution
actually has $\nu =3.5$, the resultant corrected number of resonances will
be too large. To illustrate this point, we adapted the technique of Ref. 
\cite{Fu65} to a $\nu =3$ distribution. Applying the technique of Ref. \cite%
{Fu65} (with $c^{(1)}=1.22\times 10^{-6}$) to the data in the $350<E_{n}<700$
eV region, assuming $\nu =1$ results in a corrected average level spacing of 
$4.93\pm 0.35$ eV, which is $2.3$ standard deviations \cite{Mu81} different
from the corrected value ($5.74\pm 0.40$ eV) in the $E_{n}=0-350$ eV region.
In contrast, applying this same technique to these same data, but assuming $%
\nu =3$ results in a corrected average level spacing of $6.25\pm 0.44$ eV ($%
N_{corr}=56$), only 1.2 standard deviations from the result obtained for the 
$E_{n}=0-350$ eV region. Hence, these calculations indicate the approach we
have taken is reasonable, and further indicate that the $\Gamma _{n}^{0}$
data in the $350<E_{n}<700$ eV region are inconsistent with a PT
distribution.

As a further check on the correction for missed resonances, we applied the
technique of Ref. \cite{Ge81}, which is based on the $\Delta _{3}$
statistic. The present case is very similar to the $^{235}$U example
discussed in Ref. \cite{Ge81}, from which it can be calculated that most
likely $0_{-0}^{+5}$ $^{147}$Sm+$n$ resonances were missed for each spin
state for $E_{n}<700$ eV. Hence, the corrected number of resonances for $%
E_{n}<700$ eV from this technique is smaller than, but consistent with, the
value obtained above following the technique of Ref. \cite{Fu65}.

One problem with using the KS test is that it is nonparametric, but we have
determined parameters of the theoretical distribution from the data. In such
cases, Ref. \cite{Co80} indicates that the KS test is conservative, and
Refs. \cite{Co80,Li67} describe how to modify the KS test to make it
parametric: The test statistic remains unchanged, but different tables of
critical values are used, and these values are calculated using Monte Carlo
techniques.

It is straightforward to adapt the KS test when $<\Gamma _{n}^{0}>$ is
determined from the data. We wrote a computer program which drew $N$ (where $%
N=54$ in the present case because this was the number of observed resonances
in the $E_{n}=350-700$-eV region) random $\Gamma _{n}^{0}$ values from a PT
distribution. The average reduced neutron width for this sampled set then
was calculated, and the maximum vertical difference (the $D^{+}$ statistic)
between a PT distribution with this $<\Gamma _{n}^{0}>$ and the random
samples was calculated. The program performed this task 30000 times to
construct a distribution of $D^{+}$ values. As a check of the program, a
second set of $D^{+}$ values was obtained in the standard KS sense (without
calculating $<\Gamma _{n}^{0}>$ from the sampled data). The $P$ values
calculated using these standard $D^{+}$ values were found to agree with
those in references (e.g., Ref. \cite{Co80}). Furthermore, it was found that
there were fewer large values of the $D^{+}$ statistic when $<\Gamma
_{n}^{0}>$ was determined from the sampled data compared to the standard KS
values, verifying that the KS test is conservative. For example, in the
present case for a PT distribution having $N=61$, the $D^{+}$ value
calculated from the data was 0.2432 (in the $E_{n}=350-700$-eV region), and
the $P$ value increased from $99.883\%$ for the KS test to $99.997\%$ for
this parametric variation.

Adapting the KS test to the case where $N$ also is determined from the data
requires additional assumptions. We assumed that the resonances were spaced
according to a Wigner distribution and that the method of Ref. \cite{Fu65}
can be used to correct for missed resonances. Hence, for the $E_{n}=350-700$%
-eV region, we assumed a starting value of $D_{0}=4.92$ eV ($N=71$), and
randomly sampled level spacings from this Wigner distribution to obtain $%
N_{Theory}$ resonance energies between $350$ and $700$ eV. We then used
random sampling to obtain a set of $N_{Theory}$ reduced neutron widths from
a PT distribution. We then applied the same threshold curve determined from
the data to remove those $\Gamma _{n}^{0}$ values which were below
threshold, resulting in $N_{Obs}$ resonances with averaged reduced neutron
width $<\Gamma _{n}^{0}>_{Obs}$. Subsequently, the method of Ref. \cite{Fu65}
was used to obtain corrected $N_{Cor}$ and $<\Gamma _{n}^{0}>_{Cor}$ values.
The PT distribution with these corrected parameters was compared to the
sampled data to obtain the $D^{+}$ value for this sample. Reduced neutron
widths below the maximum threshold for the correction technique of Ref. \cite%
{Fu65} were excluded from this calculation. This procedure was repeated
30000 times to construct the distribution of $D^{+}$ values. These
calculations revealed that when both $N$ and $<\Gamma _{n}^{0}>$ are
determined from the data, there are even fewer large $D^{+}$ values than in
either the standard KS case or the case where $<\Gamma _{n}^{0}>$ alone is
determined from the data. For example, if a PT distribution having $N=71$ is
compared to the data in the $E_{n}=350-700$-eV region, $D^{+}=0.1677$ is
obtained, for which the standard KS test yields $P=96.00\%$. In contrast,
this second modified KS test yields $P=99.98\%$ in this case.

In addition to demonstrating that the data in the $E_{n}=350-700$-eV region
are inconsistent with a PT\ distribution to high confidence, the above tests
also illustrate that this conclusion is unaltered by assuming, within
reason, a higher threshold $\Gamma _{n}^{0}$ value or more missing
resonances (than applying the method of Ref. \cite{Fu65} to the data for $%
E_{n}<350$ eV yields). For example, the final version of the "parametric" KS
test described above assumes that 10 more resonances were missed (17 versus
7) in the $E_{n}=350-700$-eV region.

\subsection{Discussion}

We have employed the same published techniques that have been used to
demonstrate the validity of the PT distribution for reduced neutron widths
to show that the PT distribution is inconsistent with the current data for $%
350<E_{n}<700$ eV to high confidence. This conclusion is in contrast with
Ref. \cite{Ko2004} where it was found that the reduced neutron width
distributions agreed fairly well with PT distributions. However, our new
DANCE data show that many of the spin assignments used in Ref. \cite{Ko2004}
as well as the relative number of $J=3$ to $J=4$ resonances assumed
(according to $2J+1$) in that reference were incorrect.

Similar deviations from a PT distribution have been reported for $^{232}$Th 
\cite{Ri69,Fo71,Fo71a,Ra72}, as well as for five odd-A nuclides ($^{151}$Sm, 
$^{163}$Dy, $^{167}$Er, $^{175}$Lu, and $^{177}$Hf) \cite{Ca76a} for which
the $\Delta _{3}$ statistic indicated that very few resonances had been
missed.

It is interesting to note that the reduced-neutron-width distribution for $%
^{232}$Th changes shape in a manner similar to what we have found for $^{147}
$Sm; from having $\nu \geq 2$ for one energy range ($E_{n}\lesssim 400$ eV) 
\cite{Ri69,Fo71,Fo71a,Ra72}, to being consistent with PT for another energy
range ($E_{n}\lesssim 2000$ eV) \cite{Ra72,Ca94}. It also is interesting to
note that the deviation from a PT distribution for $^{147}$Sm occurs at the
same energy where an anomaly in the $\alpha $ strength function ratio has
been reported \cite{Ko2004}. Finally, it may be noteworthy that all seven of
the reported deviations from PT discussed above are limited to relatively
low energies, $E_{n,max}\approx 100-700$ eV and nuclides in which
deformation may be important. Perhaps all these effects can be explained by
the same theory.

In the early days of neutron width measurements, an exponential distribution
($\nu =2$) seemed to be favored \cite{Hu55} for the reduced neutron widths.
Subsequently it was shown \cite{Po56}, however, that a PT distribution
fitted the data better. In addition to fitting the data better, plausible
arguments were put forward to explain why the underlying physics should lead
to a PT distribution. The assumptions that expansion coefficients of the
compound nuclear wave function follow a Gaussian distribution with zero
mean, that these coefficients are real (because, due to time-reversal
invariance, the reduced width amplitudes have been shown to be real \cite%
{Wi47}), and that neutron scattering is a single-channel process at these
energies, leads to the PT distribution \cite{Ly68}. Consequently, if one or
more of these conditions does not hold the result may be a width
distribution different from PT.

For example, the existence of additional channels results in $\nu $ values
greater than one. It is well known, for example, that the distribution of
total radiation widths following neutron capture is described by a $\chi
^{2} $ distribution with many degrees of freedom by virtue of the many
different possible $\gamma $-ray channels from the capturing state. However,
the lowest-lying excited state of $^{147}$Sm is at $E_{x}=121$ keV. So,
there are no known neutron channels in addition to the elastic one in the
energy range of our analysis. Furthermore, the technique used (transmission
measurement) should yield neutron widths that are fairly insensitive to
inelastic channels.

Another way of adding an additional effective channel might be through a
non-statistical nuclear structure effect such as a doorway state. It is
interesting that a (parity) doorway model has been proposed to explain the
so-called sign effect \cite{St98} in parity-violating asymmetries for 
\textit{p}-wave $^{232}$Th+\textit{n} resonances, which occurs at about the
same energy as the reported \cite{Ri69,Fo71,Fo71a,Ra72} deviation from a PT
distribution for the neutron widths in this nuclide. It was expected (based
on arguments similar to those leading to the PT distribution) that the signs
of these parity-violating asymmetries would be random. However, all ten
measured asymmetries for resonances below 250 eV had the same sign. Models
proposed to explain this sign effect are based on either distant \cite%
{Bo92,Au92,Au92a,Fl92,Ko92,Ca93a,Ca95,Le92,Au94}\ or nearby \cite%
{Au95,Fl95,Au96,De96,Hu95,Fe96,Lo96} (parity) doorway states. Perhaps the
same type of model could be invoked to explain the observed deviations in
the neutron width distributions from the expected PT shape, while at the
same time these deviations might provide some clue to the physical origins
of the doorway. The doorway might produce deviations from the PT shape by
effectively providing a second channel. In addition, it is interesting to
note the local-doorway model of Ref. \cite{Au95a} is associated with the
known octupole deformation of $^{233}$Th. Deformation also is known to be
significant in the $^{148}$Sm region \cite{Ah93}, and because deformation
could have a large effect on $\alpha $ decay, it is possible that the same
type of model might also explain the strange behavior of the $\alpha $
strength function ratio \cite{Ko2004}. There are at least two arguments
against a doorway explanation for the observed effects in $^{147}$Sm+\textit{%
n} resonances as well as the observed deviation of the $^{232}$Th+\textit{n}
neutron-width distribution from the expected PT distribution. First, the
observed effects are much narrower than expected for a doorway state.
Second, doorways having such large effects on the neutron-width
distributions presumably also should be visible (as large steps) in
strength-function plots such as those shown in the bottom part of Fig. \ref%
{D0AndS0}. However, there are no such effects visible in this figure nor in
the corresponding plot for $^{232}$Th+\textit{n} \cite{Ra72}.

Deviations from a PT distribution also may be caused by forms of symmetry
breaking. For example, isospin-symmetry breaking has been put forward \cite%
{Hu2000} as an explanation for differences between reduced-width data and a
PT distribution. However, the distributions resulting from these kinds of
symmetry breaking are expected to be superpositions of two PT distributions
rather than a $\chi ^{2}$ distribution with $\nu >1$ as observed herein.

\begin{figure}[tbp]
\includegraphics*[width=85mm,keepaspectratio]{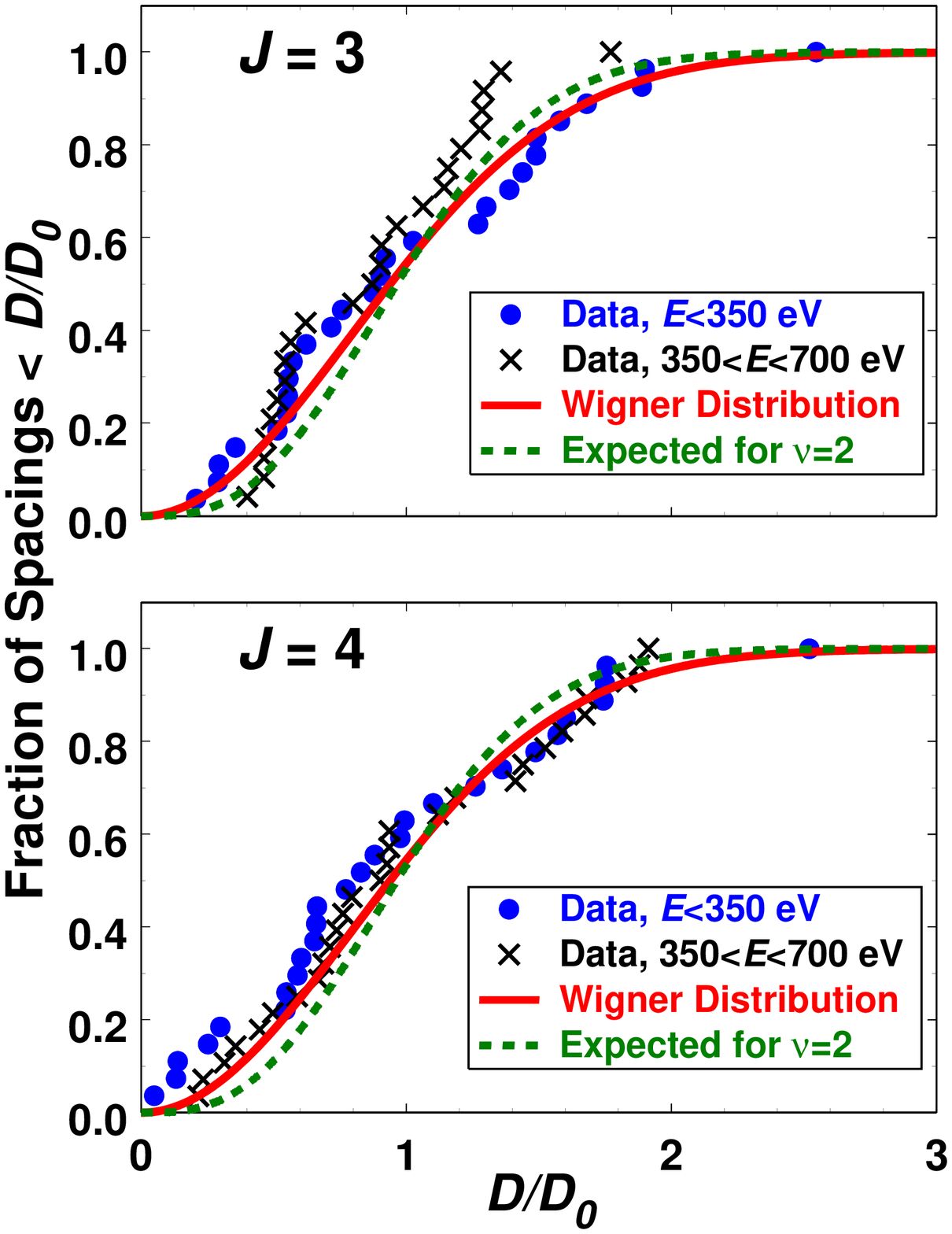}%
\caption{(Color online) Integral level-spacing distributions for $J=3$ (top)
and $4$ (bottom) resonances for two energy regions. Blue circles and black
X's depict the data for resonances below 350 eV and for 350-700 eV,
respectively. The solid red curves show the expected Wigner distributions
and the dashed green curves show the expected spacing distributions
corresponding to $\protect\chi ^{2}$ distributions for the reduced neutron
widths with two degrees of freedom.}
\label{J3And4Space2EReg}
\end{figure}

Other forms of symmetry breaking can lead to width distributions having $\nu
\geq 2$. For example, time-reversal invariance violation (TRIV) implies
compound nuclear expansion coefficients that are complex, and hence a second
degree of freedom and therefore a $\chi ^{2}$ distribution having $\nu =2$
for the neutron widths. This extra degree of freedom also should effect the
level-spacing distribution \cite{Gu98}, leading to fewer small spacings than
a Wigner distribution. Unfortunately, these effects in the level-spacing
distribution appear to be too small to observe in the present case. Spacing
distributions for both $J=3$ and $4$ for the two different energy regions
are shown in Fig. \ref{J3And4Space2EReg}. Also shown are the expected
spacing distributions corresponding to PT (Wigner distribution, or Gaussian
Orthogonal Ensemble, GOE) and $\nu =2$ (the so-called Gaussian Unitary
Ensemble, GUE) distributions for the reduced neutron widths. There is no
significant difference between the two measured distributions for $J=4$ and
the data are consistent with either theoretical distribution. Although there
is some difference between the measured distributions for the two energy
regions in the $J=3$ case, given the small number of resonances in each
region, this difference cannot be used to rule out either theoretical
distribution at a reasonable confidence level. Results from KS tests of the
various distributions are summarized in Table \ref{KSTestsTable}.

\begin{table}[tbp] \centering%
\caption{Results of Standard Kolmogorov-Smirnov Tests
\label{KSTestsTable}}%
\begin{tabular}{ccccccc}
\hline\hline
Quantity & Distribution & $J$ & $\Delta E$ (eV) & $Max$ & $N_{OBS}$ & $P$
(\%) \\ \hline
$\Gamma _{n}^{0}$ & PT & 3+4 & 0-350 & 0.0919 & 56 & 63.40 \\ 
$\Gamma _{n}^{0}$ & $\chi ^{2}$ with $\nu =2$ & 3+4 & 0-350 & 0.2435 & 56 & 
99.90 \\ 
$\Gamma _{n}^{0}$ & $\chi ^{2}$ with $\nu =3.5$ & 3+4 & 0-350 & 0.4075 & 56
& 100.00 \\ 
$\Gamma _{n}^{0}$ & PT & 3+4 & 350-700 & 0.2432 & 54 & 99.87 \\ 
$\Gamma _{n}^{0}$ & $\chi ^{2}$ with $\nu =2$ & 3+4 & 350-700 & 0.1167 & 54
& 78.72 \\ 
$\Gamma _{n}^{0}$ & $\chi ^{2}$ with $\nu =3.5$ & 3+4 & 350-700 & 0.0667 & 54
& 40.75 \\ 
$D_{0}$ & GOE & 3 & 0-350 & 0.1261 & 27 & 60.89 \\ 
$D_{0}$ & GUE & 3 & 0-350 & 0.1753 & 27 & 83.04 \\ 
$D_{0}$ & GOE & 4 & 0-350 & 0.1522 & 27 & 70.04 \\ 
$D_{0}$ & GUE & 4 & 0-350 & 0.2166 & 27 & 93.18 \\ 
$D_{0}$ & GOE & 3 & 350-700 & 0.1944 & 24 & 85.67 \\ 
$D_{0}$ & GUE & 3 & 350-700 & 0.2224 & 24 & 92.02 \\ 
$D_{0}$ & GOE & 4 & 350-700 & 0.1122 & 28 & 53.98 \\ 
$D_{0}$ & GUE & 4 & 350-700 & 0.1548 & 28 & 76.35 \\ 
$D_{0}$ & GOE & 3+4 & 0-350 & 0.0996 & 55 & 68.52 \\ 
$D_{0}$ & GUE & 3+4 & 0-350 & 0.0920 & 55 & 62.86 \\ 
$D_{0}$ & GOE & 3+4 & 350-700 & 0.1107 & 53 & 74.61 \\ 
$D_{0}$ & GUE & 3+4 & 350-700 & 0.0922 & 53 & 61.73 \\ \hline
\end{tabular}%
\end{table}%

\begin{figure}[tbp]
\includegraphics*[width=85mm,keepaspectratio]{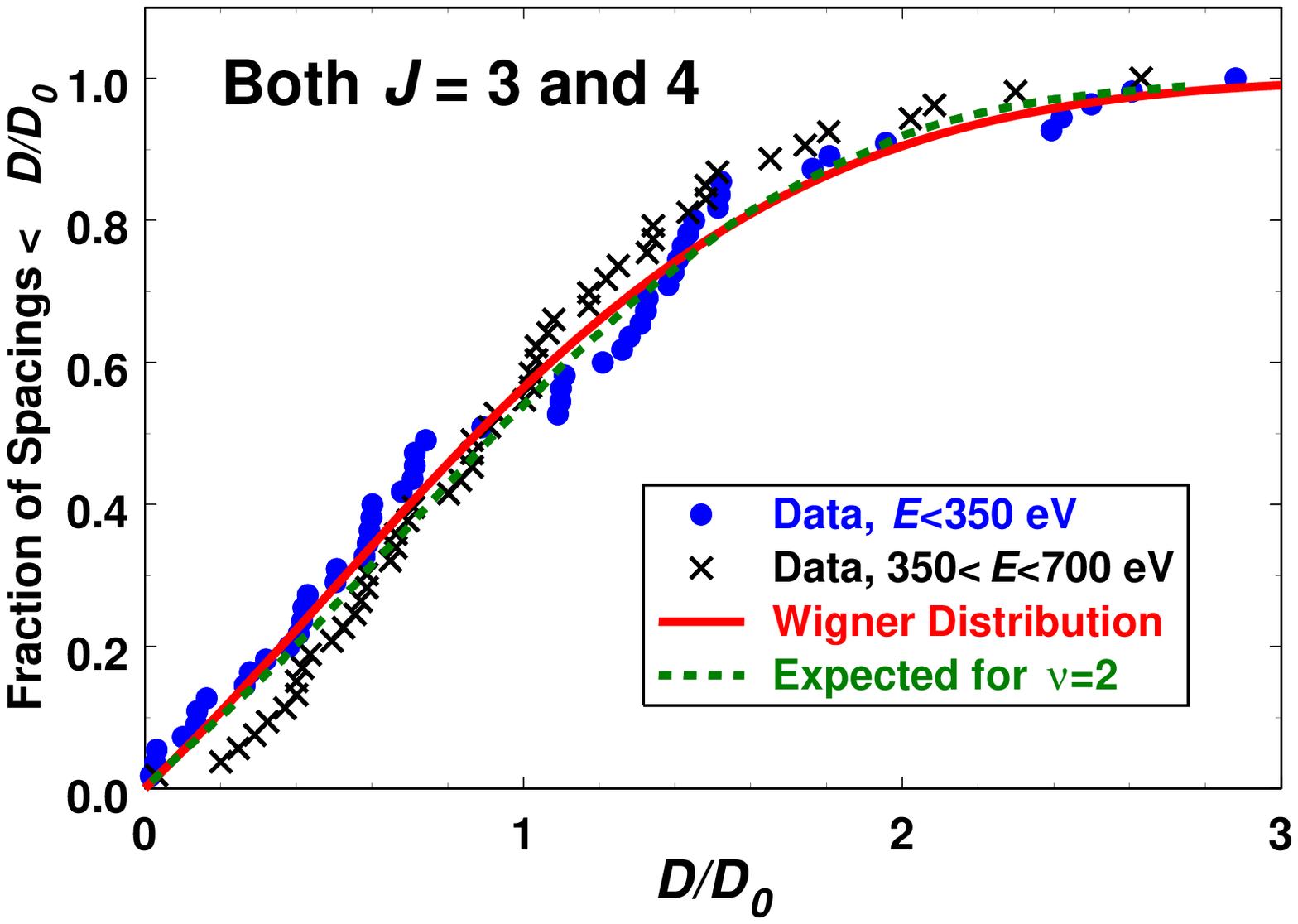}%
\caption{(Color online) Integral level-spacing distributions for combined $%
J=3$ and $4$ resonances for two energy regions. Blue circles and black X's
depict the data for resonances below 350 eV and for 350-700 eV,
respectively. The solid red curve shows the expected Wigner distribution and
the dashed green curve shows the expected spacing distribution corresponding
to a $\protect\chi ^{2}$ distribution for the reduced neutron widths with
two degrees of freedom.}
\label{BothJsSpace2EReg}
\end{figure}

Data for the two spins can be combined to increase the statistical
precision. However, combining the two spins also decreases the difference
between the two theoretical distributions. The net effect is that combining
the two spins does not improve the ability to distinguish between the two
theoretical distributions. This is shown in Fig. \ref{BothJsSpace2EReg}
where spacing distributions for the two spins combined are shown for the two
energy regions and compared to the two theoretical distributions. Although
there appears to be a difference in shape between the data in the two
regions, neither data set can be used to rule out either theoretical
distribution at the 95\% confidence level. Curiously, the level-spacing data
in the upper energy region for the two spins combined looks very similar to
a Wigner distribution for a single spin.

\section{Summary and Conclusions}

We have used information contained in multiplicity distributions of $\gamma $
rays following neutron capture to assign spins of $^{147}$Sm$+n$ resonances.
We have shown that the DANCE detector at LANSCE is an excellent apparatus
for this application. We have devised a new technique for using the measured
multiplicity information to discern resonance spins. We have demonstrated
that this new technique is superior to using the average multiplicity for
assigning spins to closely-spaced resonances. Spins were determined for 33
resonances without previous assignments and 8 firm spin assignments were
made for resonances previously having only tentative assignments. There are
several other nuclides for which this technique should be applicable and so
future measurements of this type could lead to a wealth of new resonance
parameter data.

We used these new spin assignments together with reported \cite{Mi81,Su98}\
neutron widths to determine average level spacings, neutron strength
functions, and level-spacing and reduced-neutron-width distributions for $J=3
$ and 4 resonances separately. Our analysis shows that there are very few
missing resonances below $E_{n}=700$ eV. Furthermore, using the same
techniques that have been used to correct for missed resonances and to
demonstrate the validity of the PT distribution for reduced neutron widths,
we have shown that the present data are inconsistent with PT. Specifically,
the reduced neutron width distribution changes shape near $E_{n}=350$ eV,
from being consistent with PT below this energy to being inconsistent with
PT for the next 350 eV. This change occurs at the same energy as a
previously reported \cite{Ko2004} anomaly in the $\alpha $ strength-function
ratio for $^{147}Sm(n,\alpha )$ resonances. A similar unexplained deviation
from PT was reported for neutron resonances in $^{232}$Th \cite%
{Ri69,Fo71,Fo71a,Ra72} and five odd-A nuclides \cite{Ca76} at about the same
energy. We have discussed several possible explanations for these observed
non-statistical effects. Of the considered explanations (a
previously-unknown low-lying excited state in $^{147}$Sm, a doorway state,
and TRIV) only TRIV is consistent with, but by no means proved by, the data.
Indeed we know of no physical explanation why TRIV would be manifested in
these nuclides at this energy at such levels. It seems more likely that an
unknown nuclear structure effect, perhaps one related to deformation, is
responsible for the reported anomalies \cite{Ko2004,Ri69,Fo71,Fo71a,Ra72}.
Finally, with current techniques it should be possible to significantly
improve both the accuracy and sensitivity of the previous experiment on
which the present $^{147}$Sm neutron widths are based \cite{Mi81}.
Therefore, it could be worthwhile to make new high resolution and high
sensitivity neutron capture and total cross section measurements on $^{147}$%
Sm.

\begin{acknowledgments}
We would like to thank J. D. Bowman, J. A. Harvey, G. E. Mitchell, and T. F.
Papenbrock for fruitful discussions, and the Referees for helpful
suggestions. This work was supported in part by the U.S. Department of
Energy under Contract No. DE-AC05-00OR22725 with UT-Battelle, LLC. This work
has benefited from the use of the LANSCE facility at Los Alamos National
Laboratory which was funded by the U.S. Department of Energy and operated by
the University of California under Contract W-7405-ENG-36.
\end{acknowledgments}

%\bibliographystyle{prsty}
%\bibliography{ACOMPAT,pauls}

\newif\ifabfull\abfulltrue

\end{document}